\documentclass[preprint,showpacs,amsmath,amssymb]{revtex4}

\usepackage{graphicx}
\usepackage{dcolumn}
\usepackage{bm}

\begin{document}

\title{ENERGY THRESHOLDS OF STABILITY OF THREE-PARTICLE SYSTEMS}

\author{I.V. Simenog$^{1,2}$, YU.M. Bidasyuk$^{1,2}$, B.E. Grinyuk$^1$, M.V. Kuzmenko$^1$}

\affiliation{$^{1}$Bogolyubov Institute for Theoretical Physics of
the NAS of Ukraine,  Metrolohichna Str., 14b, Kyiv-143, 03143,
Ukraine} \affiliation{$^{2}$Taras Shevchenko Kyiv National
University, 6, Academician Glushkov Prosp., Kyiv 03127}

\date{\today}

\begin{abstract}
We have studied the general properties of the energy thresholds of
stability for a three-particle system with short-range
interaction. A wide region of the interaction constants and
various ratios of the masses of particles are considered. The
specific effects characteristic of the near-threshold stationary
energy levels of three particles are revealed. The asymptotic
estimates are obtained for the thresholds at some limiting cases,
and the high-precision variational calculations of the thresholds
for various values of the interaction constants and the masses of
particles are carried out.
\end{abstract}
 \maketitle

\section{Introduction}

The studies of the quantum systems of three particles of different
nature, which are performed within various theoretical approaches,
remain to be actual for a long period (see survey \cite{R1}). This
is related to both the presence of nontrivial effects, which
appear in this simplest many-particle system, such as the famous
Efimov effect \cite{R2}, and the importance of the theoretical
consideration of real three-particle and three-cluster systems of
different nature, e.g., nuclei with three nucleons or with three
clusters, hypothetical systems of the type of trineutrons,
molecular trimers, etc. Important and insufficiently studied are
the fine near-threshold effects in the systems of three particles.
The present work is devoted to the analysis of the general
properties of these effects and the thresholds of stability for
three-particle systems. The study of properties of the
three-particle thresholds is executed qualitatively on the basis
of the analysis of the asymptotic estimates, and also using the
high-precision variational calculations with the use of optimized
Gaussian bases. This approach allows us to investigate, with a
high controlled accuracy, even such fine effects as the Efimov
effect, as well as the structure functions of these near-threshold
levels \cite{R3} that are characterized by a very small energy.

\section{Statement of the Problem}

In the present work, we consider a system of three particles, among which
two particles are identical (we set their masses $m_1=m_2=1$ without loss of
generality), and the third one can differ from them by the mass and
the pairwise interaction potential. We write the Hamiltonian of such a system in the case of
pairwise interactions in the form (in the system of units with $\hbar=
1$)
\begin{equation}
\hat H =  - {\frac{1}{2}}\Delta_1  - {\frac{1}{2}}\Delta_2 -
{\frac{1}{2m}}\Delta_3  + V(r_{12} ) + U(r_{13} ) + U(r_{23} ).
\label{E1}
\end{equation}
Let the intensities of two-particle interaction potentials be
defined by the dimensionless interaction constants $g$ (for the
pair of identical particles ${\it (12)}$) and $\lambda$ (for the
pairs of particles ${\it (13)}$ and ${\it (23)}$) on the given
form of the interaction, where $V(r) = gv(r)$ and $U(r) = \lambda
u(r)$. We consider the potential functions $v(r)$ and $ u(r)$
mainly with positive values. Negative values of $g$ and $\lambda$
correspond to the attraction, and positive ones to the repulsion.
We will study the properties of the energy thresholds of stability
for three-particle systems, i.e. the regions of such values of the
constants $g$ and $\lambda$, at which the {\it n}-th energy level
of the three-particle system appears below the two-particle
thresholds or below zero if the two-particle subsystems are not
bound,
\[
E_n (\mathrm{three}) - E_0 (\mathrm{two}) \le 0,\;\;\;E_0
(\mathrm{two}) \le 0,\]
\begin{equation}
E_n (\mathrm{three}) < 0,\text{the coupling of two particles is
absent.} \label{E2}
\end{equation}
We will study the energy thresholds of stability for
three-particle systems with the zero total angular momentum, $L =
0$, for various ratios of the masses of particles, the form of
pairwise interaction potentials, and the symmetry of the wave
function relative to the permutations of identical particles (for
the symmetric $\varphi ^s \left( {1,2;3} \right) = \varphi ^s
\left( {2,1;3} \right)$ and antisymmetric $\varphi ^a \left(
{1,2;3} \right) =  - \varphi ^a \left( {2,1;3} \right)$ states).

On the plane ($g,\lambda$), we will present the results of studies
in the form of the diagrams of the thresholds of energy levels,
namely the thresholds of stability of the energy levels of three
particles with the zero angular momentum (Fig.\ref{fig1}). The
systematization of the significant amount of calculations and the
qualitative analysis allow us to represent the thresholds of
stability in the universal form of the diagrams of thresholds. In
Fig.\ref{fig1}, we give, for the sake of specificity, the example
of the diagrams of stability corresponding mainly to potentials
$v(r)$ and $u(r)$ which have the same Gaussian form and correspond
to almost equal masses. We emphasize that the characteristic
properties of a diagram are preserved for nonsingular short-range
potentials of more general forms. In Fig.\ref{fig1}, we show the
main peculiarities of such diagrams schematically without holding
the scale, but with the preservation of all main regularities.
Each line of the threshold of the $n$-th level (for the states
$s_n$ symmetric relative to the permutation of the identical first
and second particles, and for the antisymmetric states $a_n$)
separates the region of the existence of a bound state (the region
of stability) with the corresponding $n$-th level of three
particles from the region, where this level does not exist (this
side of the curve is shaded). Because the bound states of three
particles can appear only under the attraction between different
particles (the potential $U(r)$ with the negative constant
$\lambda)$ while the interaction between identical particles (the
potential $V(r)$) can be both attractive ($g < 0$) and repulsive
($g > 0$), the three-particle thresholds are positioned to the
left from the axis of ordinates. On the axes, we marked the points
where two-particle ground bound states appear: the $s$-state with
the orbital moment $l = 0$ ($ - \lambda _{s,\mathrm{cr}}$, $
 - g_{s,\mathrm{cr}}$) and the $p$-state with the orbital moment $l = 1$ ($
 - \lambda _{p,\mathrm{cr}}$, $- g_{p,\mathrm{cr}}$). We also show
the corresponding two-particle thresholds on the axes by lines
with different thicknesses. The inclined dashed lines separate the
regions, where the lowest threshold from two two-particle ones is
the threshold for two identical particles (due to the potential
$gv(r)$, below the dashed line) or for two different particles
(due to the potential $\lambda u(r)$, above the dashed line). The
three-particle levels exist below the lowest two-particle
threshold. The dash-dotted vertical lines in the lower part of
Fig.\ref{fig1} are the asymptotes of three-particle thresholds as
$g \to - \infty$. In this case, the lower part of the figure is
correlated with the rest parts so as it follows from direct
calculations. In the diagram of the thresholds of stability, the
bound states of the system of three particles exist, by starting
from the lines of thresholds, as a rule, towards the increase in
the intensities of the attraction (i.e. to the left and down on
the diagram). On the energy diagrams of stability, we can
distinguish eight different regions by characteristic
peculiarities of the ground and excited three-particle levels:

\noindent {region I} --- the asymptotic region, where ${- \lambda
\approx g \to \infty}$,

\noindent {region II} --- the asymptotes of thresholds as $g \to -
\infty$, $\lambda = \rm Const$,

\noindent {region III} --- the region of the infinite series of
Efimov levels near the values of the two-particle interaction
constants critical as for the appearance of the bound $s$ ---
states ${g \approx  - g_{s,{\rm cr}}}$, ${\lambda \approx -
\lambda _{s,{\rm cr}}}$; on the diagram of thresholds, the
nonmonotonous and closed curves correspond to the Efimov states
(they are marked by the words ``Efimov's effect''),

\noindent {region IV} --- the axis $\lambda$ in the case of the
absence of the interaction between identical particles ${\it
(12)}$, when $g = 0$ and the conditions of the Thomas theorem are
satisfied,

\noindent {region V} --- the region where the effect of a ``tube''
is manifested for $\lambda \approx  - \lambda _{s,{\rm cr}}$ and
arbitrary positive $g$,

\noindent {region VI} --- we indicate the characteristic behavior
of thresholds for the symmetric three-particle states in the
region ${g\approx\lambda\ll - g_{s,\mathrm{cr}}}$ with the
appearance of an acute ``wedge'' on the line of equality of the
threshold binding energies of different pairs of particles,
$\varepsilon _{12}  = \varepsilon _{13}$,

\noindent {region VII} --- the region of a ``rearrangement'' of
the thresholds of energy levels,

\noindent {region VIII} --- the region of the nonmonotonicity of
the curves for the thresholds of antisymmetric levels.

The characteristic peculiarities of the behavior of the energy
thresholds of stability of the three-particle system in different
regions on the plane $(\lambda ,g)$, the general established
phenomenon of the nonmonotonicity of thresholds, and the effect of
``traps'' are studied analytically and with the use of numerical
calculations within the Galerkin variation method with a Gaussian
basis and various high-precision optimization schemes.

\section{Asymptotics of Thresholds and the Effect of a ``Tube''}

\begin{figure}[ht]
\centering
\includegraphics[width=5.0in]{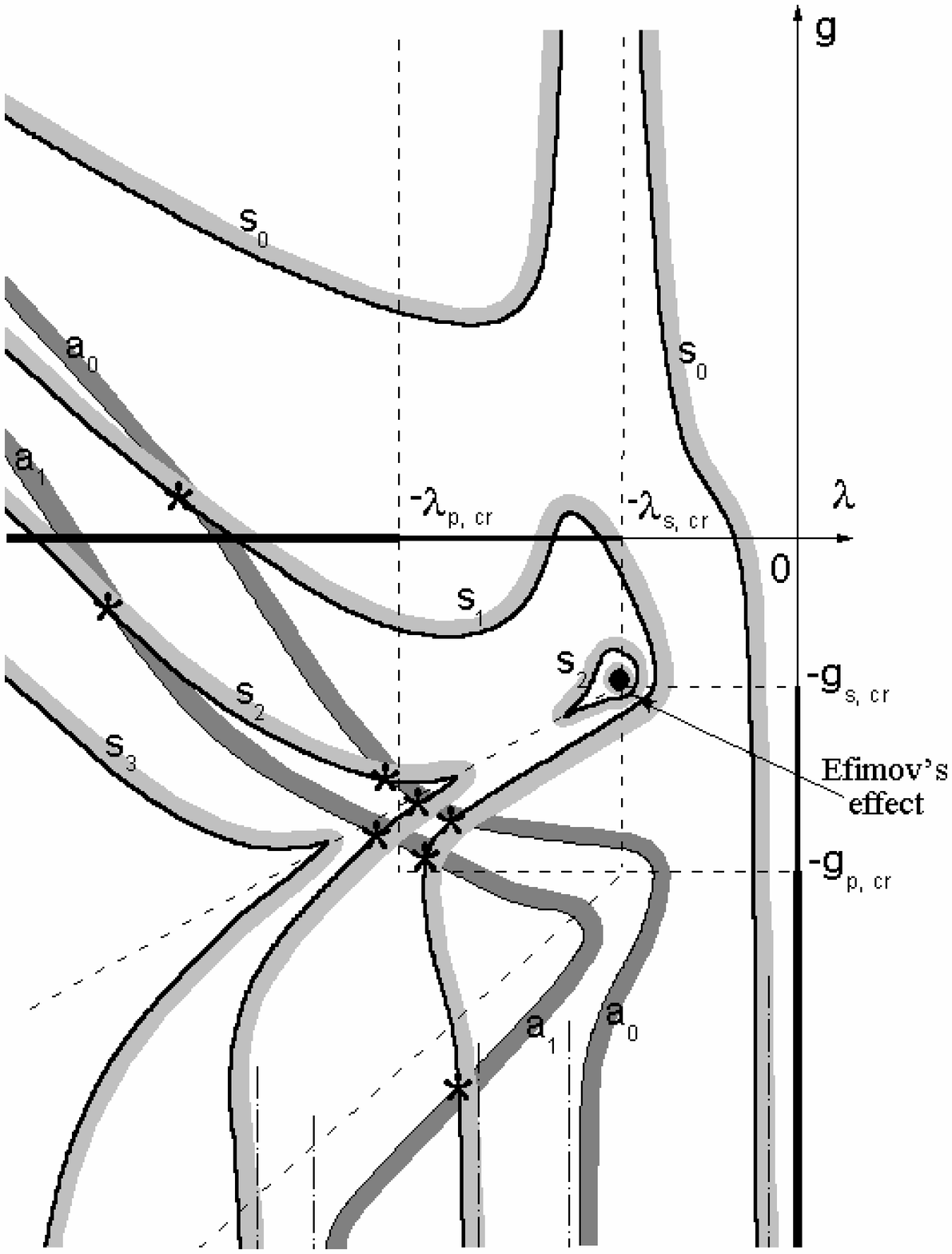}
\caption{Schematic image of the energy thresholds of stability of
a three-particle system for short-range  interaction potentials:
$s_n$ --- symmetric, $a_n$
--- antisymmetric states relative to the permutations of particles
${\it (12)}$. The asterisks indicate the intersections of the
lines of the thresholds of states with different symmetries, the
vertical dash-dotted lines in the lower part of the figure are the
asymptotes for the lines of thresholds, and the dashed inclined
lines correspond to the equality of the binding energies
$\epsilon_{12}=\epsilon_{13}$ of different pairs of particles.}
\label{fig1}
\end{figure}

Consider { region I} for the ground and excited symmetric states
on the diagram of thresholds, where $\lambda  \to  - \infty$ and
simultaneously $g \to  + \infty$ (the left upper part of the
diagram of thresholds in Fig.\ref{fig1}). We rewrite Hamiltonian
(\ref{E1}) in the center-of-mass system as
\begin{equation}
H =  - {\frac{1}{2m}}\left(1 + {\frac{m}{2}}\right)\Delta _\rho -
\Delta _r + gv(r) + \lambda u\left( {\left| {{\bm {\rho }} +
\frac{{\bf{r}}}{2}} \right|} \right) + \lambda u\left( {\left|
{{\bm {\rho }} - \frac{{\bf{r}}}{2}} \right|} \right), \label{E3}
\end{equation}
where ${\bf{r}} = {\bf{r}}_1  - {\bf{r}}_2 $, ${\bm {\rho }} =
{\bf{r}}_3  - \left( {{\bf{r}}_1  + {\bf{r}}_2 } \right)/2$ ---
Jacobi relative coordinates. Let a short-range repulsive potential
$gv(r)$ have a maximum at zero and monotonically decrease, and let
the potential $\lambda u(r)$ have a minimum value $U_0 $ ($U_0  <
0$ --- the attraction, and this occurs not obligatorily at $r =
0$). In the limit of strong coupling, the main contribution to the
ground-state energy is determined by the minimum of the full
three-particle potential well, which is positioned at $\rho _{\min
}  = 0$ and at some $r_{\min }$, minimizing the effective total
potential energy $V(r) + 2U(r/2)$. For nonsingular short-range
potentials $V$ and $U$, the threshold line is determined by the
formula
\begin{equation}
\mathop {\min }\limits_r (gv(r) + 2\lambda u(r/2)) = \mathop {\min
}\limits_x \lambda u(x), \label{E4}
\end{equation}
which determines the value of $r_{\min }$, where the minimum of
the effective attractive potential energy of three particles
$gv(r) + 2\lambda u(r/2)$ is attained, and establishes the
connection between the constants $g$ and $\lambda $ in the
considered region of the diagram of thresholds. In the simple case
of the interaction potentials $V$ and $U$ in the Gaussian form
with unit radii, which is thoroughly studied numerically, the
condition for the threshold has the form
\begin{equation}
ge^{ - r_{\min }^2 }  - 2\left| \lambda  \right|e^{ - r_{\min }^2
/4}  =  - \left| \lambda  \right|. \label{E5}
\end{equation}
From the condition of the minimum of the left-hand side of
(\ref{E5}), we get
\begin{equation}
r_{\min }  = \sqrt{\frac{4}{3}\ln \left( \frac{2g}{\left| \lambda
\right|} \right)} , \label{E6}
\end{equation}
and, with regard for (\ref{E5}),
\begin{equation}
r_{\min }^2  = 4\ln (3/2)\approx 1.62. \label{E7}
\end{equation}
Then, in the limit $\left| \lambda  \right| \to
\infty$, the asymptotic formula with regard for the next correction for the lines of the thresholds
of the ground and excited levels has finally the following form:
\begin{equation}
g =\frac{{27}}{{16}}\left( {\left| \lambda  \right| + C_{p,q}
\sqrt {2\left| \lambda  \right|}  + O(1)} \right). \label{E8}
\end{equation}
Here,
\begin{equation}
C_{p,q}  = \frac{9}{2}\sqrt {1 + \frac{1}{m}}  - \left(2p +
1\right)\frac{{3r_{\min } }}{2} - \left(2q + 3\right)\sqrt
{\left(1 + \frac{2}{m}\right)\left(\frac{3}{2} - \frac{{r_{\min
}^2 }}{4}\right)},\;\; p,q = 0,1,2,... \label{E9}
\end{equation}

We note that the quantum numbers $p = q = 0$ in (\ref{E8}) and
(\ref{E9}) correspond to the ground state. The series of states in
$p$ is due to one-dimensional small oscillations along the
coordinate $x \equiv r - r_{\min } $, whereas the series of states
in $q$ corresponds to three-dimensional oscillations along the
coordinate $\rho$ near the minimum of the potential well in the
three-particle system. Indeed, if we expand the potential energy
(in the case of the Gaussian potentials $V$ and $U$) in
Hamiltonian (\ref{E3}) averaged over angles near the minimum in
$\rho ^2 $ and in the squared deviation $x^2  \equiv \left( {r -
r_{\min } } \right)^2 $, we get, instead of (\ref{E3}), the
approximate oscillatory Hamiltonian as a function of both
coordinates
\begin{equation}
\tilde H_{\rho ,x}=- {\frac{1}{2m}}\left(1 + {\frac{m}
{2}}\right)\Delta _\rho   + \frac{2}{3}\left( {2 - \frac{4}{3}\ln
\left( {\frac{3}{2}} \right)} \right){\rm{ }}\left| \lambda
\right|\rho ^2 -  \frac{1}{{x^2 }}\frac{\partial }{{\partial
x}}x^2 \frac{\partial }{{\partial x}} + 2\ln \left( {\frac{3}{2}}
\right){\rm{ }}\left| \lambda  \right|x^2 . \label{E10}
\end{equation}
The energy states of the oscillatory  Hamiltonian (\ref{E10})
generate the lines of thresholds (\ref{E8}) and (\ref{E9}). The motion of the system of three
particles in symmetric states in approximation (\ref{E10}) is
oscillations relative to the following configuration. At the center of the system,
a particle with mass $m$ is located. On the same diameter with it, particles 1 and
2 ($\rho  = 0$) are positioned, but the wave function is spherically symmetric in angles
(the sphere diameter $r_{\min }  \approx 2\sqrt{\ln \left(3/2\right)}
  + {\rm O}\left( {1}/{\sqrt g } \right)$). The distance between
particles 1 and 2 does not depend in the main approximation on the mass
of the third particle. That is, such a configuration of the system (the third particle
is at the center between the two identical particles) occurs even in the case where the mass $m$
of the third particle is small.

The more detailed consideration of the threshold in {region I}
shows that asymptote (\ref{E8}) for the ground state is defined as
\begin{equation}
g \to \frac{{27}}{{16}}\left( {\left| \lambda  \right| + C_{0,0}
\sqrt {2\left| \lambda  \right|}  + ...} \right)\;, \label{E11}
\end{equation}
and the coefficient
\begin{equation}
C_{0,0}  =  - \frac{3}{\sqrt 2 } \left\{\sqrt {\frac{2 +
m}{m}\left( 3 - 2\ln \left( \frac{3}{2} \right) \right)} \right.
\left. + \sqrt {2\ln \left( \frac{3}{2} \right)} - 3\sqrt {\frac{1
+ m}{2m}} \right\} \label{E12}
\end{equation}
is positive at very small masses. Moreover, $C_{0,0}  \to 3\left(
{{3\left/2\right.}} - \sqrt {3 - 2\ln \left( {{3\left/2\right.}}
\right)} \right)\left/{\sqrt m }\right. \approx 0.06\left/ \sqrt m
\right.$ as $m \to 0$, is negative at greater masses and reaches a
minimum value at $m \approx 0.057$ ($C_{0,0}  \approx  - 1.39$).
At $m \to \infty $, it approaches the constant, $C_{0,0}  \approx
- 0.55$. Thus, the linear asymptote of the threshold for the
ground state in {region I} of the threshold diagram in
Fig.\ref{fig1} is reached from the top for very small masses ($m
\le \sim0.001$) and is reached from the bottom for the greater
mass of the third particle.

We also note that, in {region I} of the diagram of thresholds, the
result which is asymptotic in the coupling constant, $\left|
\lambda \right| \to \infty$, is easily generalized to other
potentials. In particular, let the interaction potential $U$
between different particles be chosen in the Gaussian form with
another interaction radius $R \ne 1$. Then we get the same
configuration of three particles (the third particle is located
between two identical ones). Moreover, instead of (\ref{E7}), we
obtain (for $R \ge 1\left/{\sqrt 2 }\right.$, when $r_{\min }^2
\ge 0$)
\begin{equation}
r_{\min }^2  = 4R^2 \ln\frac{{4R^2  - 1}}{{2R^2 }}. \label{E13}
\end{equation}
In the main approximation, we get
\begin{equation}
g = B\left( R \right) \left| \lambda  \right| + O\left( {\sqrt
{\left| \lambda  \right|} } \right)\;, \label{E14}
\end{equation}
where
\begin{equation}
B\left( R \right) = \frac{{\left( {4R^2  - 1} \right)^{4R^2  - 1}
}}{{\left( {2R^2 } \right)^{4R^2 } }}\;. \label{E15}
\end{equation}

The least value of the coefficient, $B\left( R \right) = 1$,
is reached in (\ref{E14}) at $R = {\sqrt 2 }\left/ 2 \right.$, when
$r_{\min } = 0$ and all three particles approach one another at small distances.
As $R \to \infty $, the coefficient $B\left( R \right)$ grows
indefinitely. For $R < {\sqrt 2 }\left/ 2 \right.$, all three particles are,
all the more, at small distances, where the small oscillations
of the particles relative to the equilibrium position occur. The asymptotics
of a threshold remains invariable: $g = \left| \lambda \right| + {\rm
O}\left( {\sqrt {\left| \lambda  \right|} } \right)$. Hence, in
the region of the diagram of thresholds where $g \to \infty $ and $\lambda  \to  - \infty
$, the three-particle system can possess different configurations even in the
case of the simplest interactions. Qualitatively, analogous results are
obtained for a wider class of the pairs of potentials $V$ and  $U$. For example,
for potentials in the form of exponentials, $V(r) = ge^{ - r} $ and $U(r) =
- \left| \lambda  \right|e^{ - r/R}$, we have, for $R \ge 1$,
\begin{equation}
r_{\min }  = 2R\ln\frac{{2R - 1}}{R}\;,\; B(R) =
\frac{1}{R}\left(2 - \frac{1}{R}\right)^{2R - 1}\;. \label{E16}
\end{equation}
In this case, the least value of the coefficient $B\left( R
\right) = 1$ is also attained at $R = 1$, when $r_{\min }  = 0$,
and both $r_{\min }$ and $B(R)$ grow with increase in $R$.
Moreover, for an arbitrary pair of the monotonous repulsive
potential $V(r)$ between the identical particles and the
attractive potential $U(r)$ between different particles in {region
I}, a linear dependence between the intensities of the potentials
takes place, which separates the region of stability of three
particles from the region, where the coupling is absent. If the
repulsive potential $V$ is nonmonotonous and has a sufficient
decrease at zero, and if the attractive potential $U$ has a
minimum at finite distances $r_1  > 0$, a two-cluster
configuration of the system of three particles is also possible:
in this case, the identical particles are positioned near each
other, and the third one is located at a distance $r_1 $ from
them.

All the main qualitative and analytic results concerning the
asymptotics of the thresholds of stability in {region I} are
confirmed by the high-precision systematic calculations with
two-particle potentials of the Gaussian form for a great variety
of masses within the variational approach with the use of Gaussian
bases. The separate calculations were executed for other
potentials, and they also support the general schematic
Fig.~\ref{fig1}.

We now consider the asymptotic limit of the strong coupling
between the identical first and second particles ({region II},
where $g \to - \infty $ in the lower part of the diagram of
thresholds (Fig.\ref{fig1})). In this limit, the size of subsystem
${\it (12)}$ tends to zero, and the variables in (\ref{E1}) are
separated in the cluster approach. Then the three-particle wave
functions for Hamiltonian (\ref{E3}) can be chosen as
\begin{equation}
\Psi _n ({\bm r},{\bm \rho} ) \approx \varphi _0^{(\mathrm{osc})}
(r)f_n (\rho )\;, \label{E17}
\end{equation}
where $\varphi _0^{(\mathrm{osc})} (r)$ is the ground state
oscillatory wave function of the relative coordinate ${\bm r}$ of
pair ${\it (12)}$, and the wave function of the third particle,
being in the effective field of two other particles, depends on
the coordinate ${\bm \rho}$ and is the eigenfunction of the
effective Hamiltonian reckoned from the two-particle threshold,
\begin{equation}
\label{E18} h =  - \frac{1}{{2\mu }}\Delta _\rho   +
U_{\mathrm{eff}} (\rho )\;,
\end{equation}
with the reduced mass $\mu  = 2m(m + 2)^{-1}$  and  the effective
averaged potential
\begin{equation}
\label{E19} U_{\mathrm{eff}} (\rho ) =  - \frac{2\left| \lambda
\right|}{{\pi ^{3/2} }}\int {d{\bm r}e^{ - r^2 } } u\left( {\left|
{{\bm \rho}  - {\bm r}/\left( {2\left| g \right|^{1/4} } \right)}
\right|} \right).
\end{equation}
Consider the motion of the third particle relative to the pair of
particles ${\it (12)}$ in potential (\ref{E19}) which looks, in
particular for the Gaussian form, as
\begin{equation}
U_{\mathrm{eff}} (\rho ) =  - \frac{{2\left| \lambda
\right|}}{{\left( {1 + 1/\left( {4\sqrt {\left| g \right|} }
\right)} \right)^{3/2} }}e^{ - \rho ^2 /\left( {1 + 1/\left(
{4\sqrt {\left| g \right|} } \right)} \right)}. \label{E20}
\end{equation}
Then, for the critical constants $\lambda _{\mathrm{cr}}^{(n)}$,
at which the $n$-th three-particle near-threshold level in the
three-particle system appears, we get the asymptotics as $g \to  -
\infty $:
\begin{equation}
 \lambda _{cr}^{(n)}  \approx  - \frac{{m +
2}}{{8m}}g_{s,\mathrm{cr}}^{(n)} \left(1 + \frac{1}{{8\sqrt
{\left| g \right|} }}\right)\;. \label{E21}
\end{equation}
Here, $g_{s,\mathrm{cr}}^{(n)}$ are the critical constants of the
appearance of the $n$-th level in the system of two particles with
the zero orbital moment $l = 0$ (for the Gaussian potential with
unit radius in the case of the unit masses of particles, the
critical constants of the $s$-states are as follows:
$g_{s,\mathrm{cr}}^{(0)} \equiv g_{s,\mathrm{cr}}  = 2.684005$;
$g_{s,{\rm cr}}^{(1)}  = 17.79570$; $g_{s,\mathrm{cr}}^{(2)}  =
{\rm{45}}{\rm{.57348}}$; $g_{s,\mathrm{cr}}^{(3)}  =
{\rm{85}}{\rm{.96340}}$, etc.). The expressions analogous to
(\ref{E21}) can be also obtained in the case where the subsystem
of two particles in the state with the wave function (\ref{E17})
has a nonzero orbital moment. It follows from (\ref{E21}) that the
threshold of each three-particle level (in the states symmetric in
the permutations of the pair of particles ${\it (12)}$) is
positioned always to the left from the corresponding vertical
asymptote for $\lambda _{\mathrm{cr}}^{(n)} $ (they are marked by
dash-dotted vertical lines in the lower part of Fig.\ref{fig1}).
We can show analogously that, for the states antisymmetric
relative to the permutations of the pair of particles ${\it
(12)}$, the thresholds approach, on the contrary, the
corresponding vertical asymptotes from the right side. The
high-precision calculations for different masses and potentials in
the form of a superposition of Gaussian functions confirm
completely the above-presented conclusions about the asymptotics
of thresholds for different levels as {$g \to  -
\infty $.}%

Consider {region III} in the vicinity of the critical constants of
the appearance of the two-particle ground bound $s$-states: $g \to
- g_{s,\mathrm{cr}}^{} $ and $\lambda  \to  - \lambda
_{s,\mathrm{cr}} $ in Fig.\ref{fig1}. This is the region where the
Efimov effect \cite{R2} is manifested, and the infinite series of
Efimov weakly bound levels for the system of three (generally
saying, different) particles is observed. In the diagram of
thresholds, it is seen as the infinite collection of closed
self-similar (with the universal scale for highly excited states)
convex curves are accumulated at the critical point ($ - \lambda
_{s,\mathrm{cr}}^{} ,{\rm{ }} - g_{s,\mathrm{cr}}^{} $) for three
pairs of particles interacting in a resonance way. The lines of
thresholds for the series of Efimov energy levels are considerably
stretched along the dashed inclined line (which starts from the
region of the Efimov effect), where the two-particle thresholds of
different pairs of particles in the $s$-state coincide, and are
somewhat elongated vertically upward along the $g$ axis, when only
two pairs of particles interact in a resonance way. The
fundamental Efimov effect is described in detail by V. Efimov
himself and many other researchers. Here, we should like to
emphasize only separate points.  Firstly, the infinite series of
symmetric three-particle Efimov levels with the zero angular
momentum is realized only in the limit $g \to  - g_{s,{\rm cr}} $,
$\lambda  \to  - \lambda _{s,\mathrm{cr}} $, i.e. in the case
where all three pairs of particles in singlet states are at a
resonance. Outside this region, the number of levels is always
bounded, though these levels have the certain specific properties
of properly Efimov levels especially in the case of two resonating
pairs of particles ${\it (13)}$ and ${\it (23)}$ ($\lambda \to -
\lambda _{s,\mathrm{cr}} $) even under a sufficiently large
decrease in the attraction between the particles of the third
pair, ${\it (12)}$. Secondly, by the example of the series of
Efimov levels, we have the ideal demonstration of the energy
``traps'' (see Section 5 for more details), when the strengthening
of the attraction on the diagram of thresholds leads to a
nonmonotonous variation of the number of energy levels: the
three-particle levels at first separate from the two-particle
threshold and then disappear on it. The Efimov levels exist only
in a certain interval of the coupling constants of all three pairs
of particles interacting in a resonance way. Thirdly, for
three-particle states antisymmetric in the pair of the identical
particles ${\it (12)}$, the singlet point ($ - \lambda
_{s,\mathrm{cr}}, - g_{s,\mathrm{cr}}^{} $) is nonsingular (see
Fig.\ref{fig1}, {region VIII}). At the same time, the limiting
region near the point ($ - \lambda _{s,\mathrm{cr}} , -
g_{p,\mathrm{cr}} $), where one pair ${\it (12)}$ of the identical
particles resonates in the state with the angular momentum $l =
1$, is, as if, ``attractive'' for the corresponding states. In
this region of Fig.\ref{fig1}, there exists a certain anomaly: the
curves for the thresholds of antisymmetric states are
significantly stretched towards small coupling constants along the
line where the two-particle thresholds of two different pairs of
particles in the singlet and triplet states coincide. Finally, we
note that we do not discuss the three-particle antisymmetric
states with nonzero angular momentum, where there exists the
possibility for a collapse to arise \cite{R4}, which requires a
separate consideration.

Consider the case where the interaction of one pair of particles,
${\it (12)}$ is absent (when $g = 0$). This corresponds to {
region IV} on the $\lambda $ axis in Fig.\ref{fig1}, where the
conditions of the Thomas theorem \cite{R5} are satisfied. Then,
there always exists the bound ground state of three particles in a
wider region of coupling constants $\lambda $ of two pairs of
particles as compared with an isolated pair of two particles. This
means that the three-particle threshold of the ground state for $g
= 0$ is always positioned to the right from the two-particle
critical point in the constant $\lambda $. First of all, we note
that the important common point for the Thomas and Efimov effects
is the presence of pairs of particles interacting in a resonance
way. Moreover, the regions of these effects are positioned near
each other in the diagram of thresholds (Fig.~\ref{fig1}). At the
same time, the Thomas effect is referred only to the ground state
of three particles, whereas the Efimov effect concerns the
infinite series of near-threshold weakly bound states. Secondly,
the universality of the Thomas theorem consists in that it holds
for Hamiltonian (\ref{E3}) with $g = 0$ for an arbitrary
attractive potential $\lambda u(r)$ between two pairs of particles
and an arbitrary finite mass $m$ of the third particle. Moreover,
the less the mass $m$, the wider is the region of the
manifestation of the Thomas effect relative to the interaction
constant. We note that, for small masses, the number of
three-particle levels becomes great, when the two-particle
subsystems are not bound yet. Thirdly, there occurs the effect of
a ``trap'' for the first excited state at $g = 0$ for particles
with close masses (under certain conditions, this concerns higher
excited states too), as distinct from the ground state. In this
case, the existence of an excited level depends nonmonotonically
on strengthening the attraction by variation of $\left| \lambda
\right|$. In addition, the intersection points of the lines of the
thresholds of excited levels with the abscissa axis
(Fig.\ref{fig1}) from the side of great values of $\left| \lambda
\right|$ are located to the right from the corresponding
two-particle critical values of the coupling constants, where the
binding energies of two particles are equal to zero. We emphasize
once more that the schematic Figure 1 represents all main
regularities so as they follow from the asymptotic estimates and
the high-precision calculations.

\begin{figure}[ht]
\centering
\includegraphics[width=6.in]{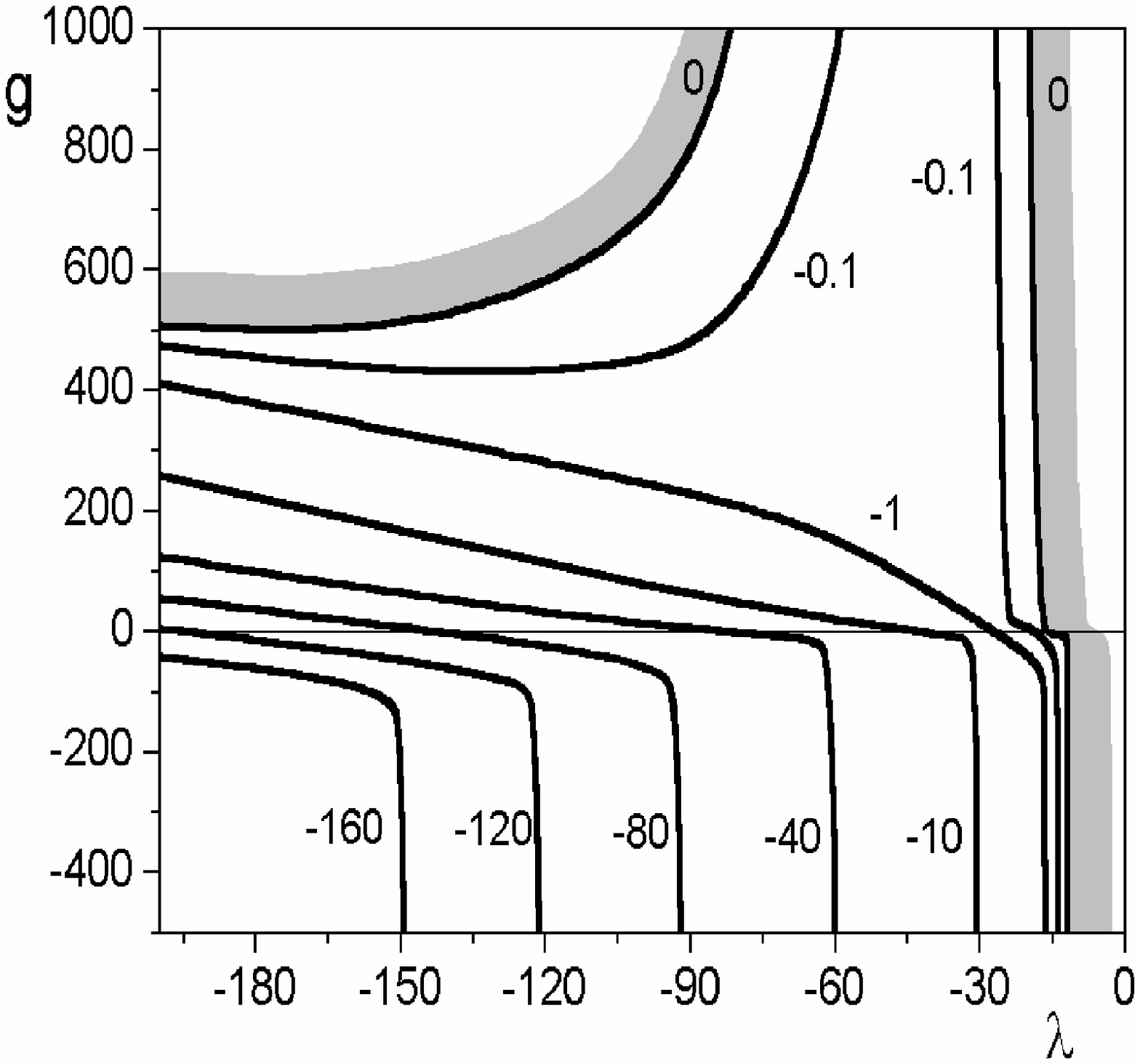}
\caption{Isolines of the ground state energy of the three-particle
system for $m=0.06$ with Gaussian potentials. At the calculated
isolines, we indicate the difference of the energies of three and
two particles.} \label{fig2}
\end{figure}

A special attention should be paid to the effect of a ``tube'' for
symmetric states at the diagram of thresholds ({region V} in the
upper part of Fig.\ref{fig1}), where the bounded region near
$\lambda \sim - \lambda _{s,\mathrm{cr}} $ contains at least one
bound state of three particles even under the unlimited increase
of the repulsion between the pair of identical particles ($g$ is
arbitrary). The effect of a ``tube'' is seen more clearly in
Fig.\ref{fig2}, where, on the real scale, we show the results of
high-precision calculations for the threshold of the ground state
and the isolines of the energy of three particles, which is
reckoned from the two-particle threshold, for the Gaussian
potentials with unit interaction radii for the mass $m = 0.06$.
The form of the energy surface in the region $g \gg 1$ and near
$\lambda \sim - 40$, indeed, reminds of a ``tube'' positioned
vertically. It is seen that the isoenergetic lines $\Delta E = \rm
const$ for the ground state of three particles with small values
of $\Delta E \approx  - 0.1$ are pulled into the ``tube'' in the
upper part of the figure for $\lambda \approx  - 40$ and great
repulsive constants $g$. The isolines do not form a ``tube''
already at $\Delta E \approx  - 1$ and greater values, but reveal
the clear asymptotic behavior characteristic of the isolines with
great values of $\Delta E$, showing a sharp change of the modes
near the abscissa axis (on the given scale). It is important that,
on the increase of the attraction between particles ${\it (13)}$
and ${\it (23)}$ (the motion along the horizontal axis in
Fig.\ref{fig2}) under a significant repulsion between the pair of
particles ${\it (12)}$, the nonmonotonicity in the number of bound
states is always observed (we call this as the effect of a
``trap'' related to the presence of the ``tube'').

The phenomenon of a ``tube'' can be explained on a qualitative
level. We will show firstly that the number of levels in a
``tube'' is always bounded, but it grows infinitely as $m \to 0$.
We use the fact \cite{R3} that the pair correlation functions
$G_{(n)} (r) \equiv \left\langle {\varphi _n } \right|\delta
({\bf{r}} - {\bf{r}}_{12} )\left| {\varphi _n } \right\rangle $ in
Efimov states $\left| {\varphi _n } \right\rangle $ for $r = 0$
for the adjacent levels satisfy the relation
\begin{equation}
G_{(n + 1)} (0)\left/
 G_{(n)} (0)\right. \to \frac{1}{{\sqrt {\Lambda
_0 } }}\;, \label{E22}
\end{equation}
where $\Lambda _0 $ is the ratio of the energies of adjacent
Efimov levels in the limit $n \to \infty $  ($\Lambda _0  =
e^{2\pi /s_o } = 515.035\ldots$ in the case of three identical
particles, and the Danilov---Minlos---Faddeev---Efimov constant
$s_0  = 1.00623$ determines the index of singularity of the
limiting equation of Skornyakov---Ter-Martirosyan in the case of
the zero interaction radius). Relation (\ref{E22}) takes place due
to the fact that, as $n \to \infty $, the spatial region, where
the pair correlation function behaves itself as $G (r)\sim r^{-2}$
\cite{R3}, is spreading. It is true from distances of the order of
the radius of forces to distances of the order of the size of the
system $\sim{\rm{ }}\left( {\sqrt {\Lambda _0 } } \right)^n R_0 $
(here, $R_0 $ is the characteristic size of the system in the
ground state, $n$ is the level number, and $\sqrt {\Lambda _0 }
\approx 22.69$ is the ratio of sizes of the system for adjacent
Efimov levels). Therefore, the pair correlation function for
sufficiently great $n$ has the following normalizing factor of
${\sim r^{-2}}$:
\begin{equation}
C_n  \to \left( \int\limits_{r_0 }^{\left( {\sqrt {\Lambda _0 } }
\right)^n R_0 }r^{-2 }d{\bf{r}} \right)^{ - 1} \to \sim{{\left(
{\sqrt {\Lambda _0 } } \right)^{-n} R_0^{-1} }}\,.
 \label{E222}
\end{equation}
At small distances, the pair correlation function is of the same
order as that at $r\sim r_0 $, therefore we obtain relation
(\ref{E22}).

In order to approach the region of the ``tube'' from the region,
where the Efimov effect manifests itself, we consider some
additional short-range repulsion $W(r_{12} )$ between the pair of
particles ${\it (12)}$ which is taken as a perturbation of the
zero Hamiltonian, for which the Efimov effect is observed. The
greater the number of a level, the greater is the size of the
three-particle system, which reminds of a spherical concentric
``halo'' \cite{R3} with the ratio of the radii of the adjacent
spheres $\sim\sqrt {\Lambda _0 } $. The greater the size of a
system of particles with short-range interaction, the better are
satisfied the conditions for the ``gas'' approach \cite{R6} taking
into account a contribution of the short-range repulsion
$W(r_{12})$ to the energy. In the principal approximation in the
``gas'' parameter, a shift of the $n$-th level is determined by
the two-particle $T$-matrix:
\begin{equation}
\Delta E_n  = \left\langle {\varphi _n } \right|T_{12} \left|
{\varphi _n } \right\rangle  \to \frac{{4\pi }}{m}a  G_{12,n}
(0)\;. \label{E23}
\end{equation}
Here, $T_{12} $ is the two-particle $T$-matrix defined for the
potential $W(r_{12} )$, $a$ is the scattering length by the
potential $W(r_{12} )$ (for the repulsive potentials, $a > 0$).
These quantities are finite for short-range repulsive potentials
of the given form in the limit of the infinitely great repulsion,
$g \to \infty $. In this case, the effective radius and other
low-energy parameters which are present in the next terms of the
expansion of the energy in the ``gas'' parameter are also finite.
These conclusions allow us to consider the most wide region in the
constant $g$ in the diagram of thresholds from the region of the
Efimov effect to the region of the ``tube'' with $g \to \infty $,
if the summary potential $ - g_{s,\mathrm{cr}} u(r_{12} ) +
W(r_{12} )$ is replaced by $gu(r_{12})$. Hence, if the intensity
of the repulsive potential increases, $g \to \infty $, the energy
shift $\Delta E_n $ obtained due to the repulsion has a finite
limit. Moreover, the greater the number $n$ of an excited level,
the more exact is the relation $\Delta E_n \to \sim a G_{(n)}
(0)\sim a { {\left( {\sqrt {\Lambda _0 } } \right)^{-n} }}$. Let
$\varepsilon _0 $ be the quantity of the order of the binding
energy of the ground state of three particles for the critical
two-particle constant. With regard for the energy $\sim{ -
\varepsilon _0 } \Lambda _0^{-n}$ of the $n$-th level of the
``zero'' Hamiltonian near the critical constants (near the Efimov
region), the total energy of the $n$-th level in the region of the
``tube'' looks as
\begin{equation}
E_n  = E_n^{(0)}  + \Delta E_n  \approx  - \varepsilon _0 {\Lambda
_0^{-n }}
 + B_0 a \Lambda _0 ^{-n/2}
\;, \label{E24}
\end{equation}
and $B_0  > 0$. Since $\Lambda _0 \gg 1$ (for $m \approx 1)$, the
energy $E_n $ becomes positive with increase in $n$ for an
arbitrary small repulsion which is characterized by the scattering
length $a$. This clarifies the above-mentioned conclusion about
the finiteness of the number of levels under a deviation from the
accumulation point of Efimov levels by the constant $g$. That is,
for the infinite number of levels to exist, it is necessary that
the interaction of all three pairs of particles be resonant, $g
\to  - g_{s,\mathrm{cr}} $ and $\lambda \to  - \lambda
_{s,\mathrm{cr}} $ simultaneously. All the more, the number of
levels becomes finite, when the repulsion $gu(r_{12} )$ reaches a
significant intensity (in this case, the $T$-matrix is bounded,
and the length $a$ tends to a finite limit of the order of the
radius of forces). At the same time, the arbitrariness of the
repulsion intensity for the pair ${\it (12)}$, the finiteness of
the $t$-matrix, and the conditions for the resonance and hence the
appearance of the effective long-range interaction for two pairs
of particles do not contradict the possibility for the bound
states of three particles to exist. The presence of at least one
bound state of three particles in the region under study is
confirmed by the numerical calculations for different forms of
potentials and for different masses. We note that the ``tube''
(see Figs.\ref{fig1} and \ref{fig2}) is asymmetrically positioned
relative to the vertical asymptote $\lambda =  - \lambda _{s,{\rm
cr}} $ (the resonance region for two pairs of particles). To a
great degree, it is positioned to the left from the asymptote,
where the attractive constant $\lambda $ is greater by modulus.
But, with the further increase in the attractive constant $\lambda
$ by modulus, we come away from the resonance region. And at
sufficiently large repulsive constants $g$, the three-particle
level can disappear, arising again only in the limit of a strong
coupling (see Figs. \ref{fig1} and \ref{fig2}). We emphasize once
more that all main qualitative conclusions of this section are
confirmed by the numerical calculations.

\begin{figure}[ht]
\centering
\includegraphics[width=4.5in]{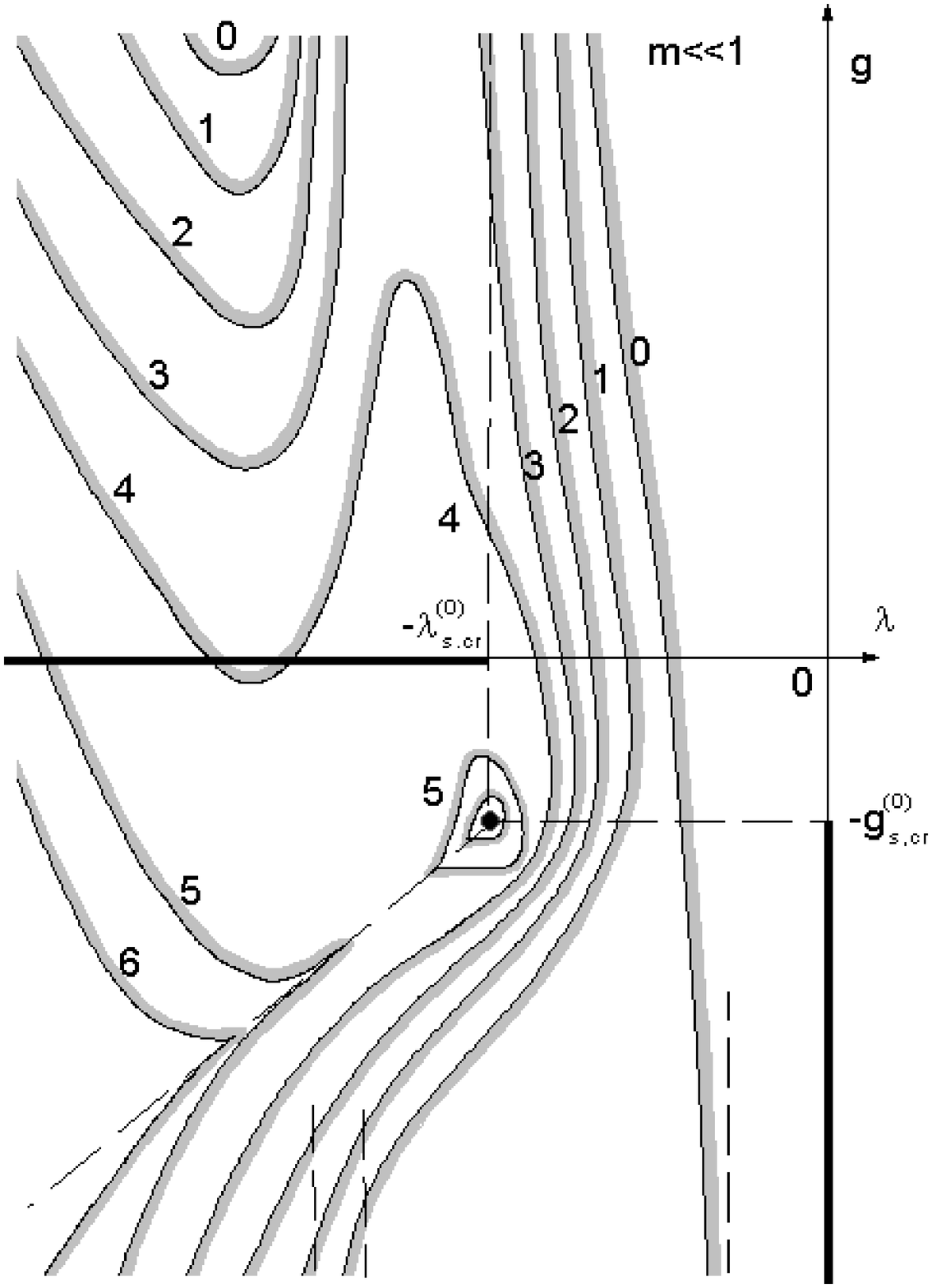}
\caption{Diagram of thresholds (schematically) in the case of a
small mass of the third particle. The numbers numerate the states
symmetric in the particles ${\it (12)}$ permutations, the other
designations are the same as in Fig.\ref{fig1}.} \label{fig3}
\end{figure}

The interesting universal phenomenon of a ``wedge'' ({region VI}
in Fig.\ref{fig1} and, respectively, in Figs. \ref{fig3} and
\ref{fig4}) is observed on the line where the binding energies of
two different pairs of particles are equal, $\varepsilon _{12}  =
\varepsilon _{13}$ (on the dashed inclined lines). For the sake of
specificity, we consider Fig.\ref{fig3}. As above, we have
correctly preserved all the regularities, which are obtained in
the calculations, though the scale of the figure is rather
arbitrary. The effect of a ``wedge'' consists in the following. If
we move along the dashed line (from the left to the right) from
the region of a strong coupling, $\left| g \right| \approx \left|
\lambda \right| \gg 1$, to the side where the attraction
decreases, then the threshold line of the $n$-th symmetric state
of three particles (in Fig.\ref{fig3}, they are the lines of the
6-th and 5-th excited states) is finished by a sharp cusp in the
form of a ``wedge'' oriented from the left to the right. The lines
of thresholds approach the dashed inclined line from the top and
from the bottom with different slopes, and the derivative of the
threshold line undergoes a break here. The ``wedge'' arises not
due to the nonanalyticity, but as a consequence of the competition
of thresholds. On the further motion to the right (within
Fig.\ref{fig3}), firstly the 6-th excited state and then the 5-th
one stop to exist. But with the decrease of the attraction along
the dashed inclined line, where the energies of the two-particle
thresholds are equal, the 5-th excited state appears again, while
we approach the Efimov region. Its threshold line has the form of
a ``wedge'' oriented from the right to the left. For higher
excited states, the effect of a ``wedge'' holds analogously.
Inside the closed region of the thresholds of Efimov states, the
bulbous curves of thresholds have a cusp in the form of a
``wedge'' on the dashed inclined line. We note that, for
antisymmetric states, no similar regularities are observed.

\section{ Dependence of the Three-Particle Thresholds on the Mass}

Firstly, we consider the case of great values of the mass $m$ of
the third particle. If the mass is infinite (the three-particle
model passes into a model with two particles in the short-range
field of a fixed center), then it is obvious from (\ref{E1}) that,
in the absence of the potential $V(r_{12} )$, the problem of three
particles is equivalent to that of two independent particles in
the field of the attractive center $U(r_i )$. The ground state of
such a system arises under the same conditions as those for each
of the particles in the field of a center:
\begin{equation}
\lambda _{0,\mathrm{cr}}  =  - \frac{1}{2}g_{s,\mathrm{cr}}^{(0)}
\;. \label{E25}
\end{equation}
In the general case of arbitrary masses, $\lambda
_{0,\mathrm{cr}}=-(1+ {1}/{m})g_{s,\mathrm{cr}}^{(0)}/2$), and,
for the Gaussian potential with unit radius, we get $\lambda
_{0,\mathrm{cr}}  =  - 1.342$. We emphasize that, for the infinite
mass of the third particle in the case where $g = 0$, the Thomas
theorem becomes trivial, and the three-particle state exists, by
beginning exactly from the two-particle critical constant. For the
excited states in the case of the infinite mass $m = \infty $, the
couplings for three and two particles also appear at the same
points on the $\lambda $ axis. In addition, the energies
monotonically increase with $\left| \lambda  \right|$. Generally,
for $m = \infty $, the thresholds of stability are transformed
only slightly on the whole plane $(\lambda ,g)$ as compared with
the case where $m\sim1$, though, in the region of the ``tube'',
there remains only the ground state which differs cardinally from
the excited states for this reason. As $m = \infty $, the whole
``tube'' is positioned to the left from the asymptote $\lambda  =
- \lambda _{0,\mathrm{cr}} $. Moreover, the right edge of the
threshold of a three-particle state crosses the abscissa axis at
the resonance point of two pairs of particles $\lambda  =  -
\lambda _{s,\mathrm{cr}} $ and rises sharply (almost vertically)
upward. In the resonance region of all three pairs of particles
($\lambda \to - g_{s,\mathrm{cr}}/2,\; g \to  - g_{s,\mathrm{cr}}$
), the conditions for the appearance of the infinite Efimov
spectrum, which possesses the universal properties by beginning
already from the second excited level, are realized in the
standard way. In other regions in the diagram of the thresholds of
stability in the case $m = \infty $, the general peculiarities
seen in Fig.\ref{fig1} are preserved as well.

We now consider the other limiting case of very small mass  $m \to
0$ (a two-center model of three particles). In the region where $g
> - g_{s,\mathrm{cr}}$, heavy particles ${\it (12)}$ are not coupled in the
absence of the third light particle, and $\lambda  \ge  - (1 +
{1}/{m})g_{s,\mathrm{cr}}/2$, as well as in the case where there
exists the coupling in the pair of heavy and light particles (pair
${\it (13)}$ or ${\it (23)}$), we can adiabatically separate the
variables. If we make averaging over the fast movement of the
light particle (in the state $\varphi $), then we get the equation
for the wave function of the coordinate of the relative motion of
particles ${\it (12)}$
\begin{equation}
\{  - \Delta _x  + mg_{12} \exp ( - mx^2 ) + W(x)\} \Phi (x) =
\varepsilon _{(12)} \Phi (x) \label{E26}
\end{equation}
(in the case of the Gaussian potentials with unit radii) with the
additional potential
\[
W(x) \!\equiv\! \int \!d{\bm{\rho }}\left| \varphi  \right|^2
\left\{\frac{\left(1\! +\! m\right)}{2}g\left( \exp \left( \!-
\left(\bm{\rho } \!-\! \frac{\sqrt m}{ 2} \bf{x}\right)^2 \right)
\right.\right. + \!\left.\exp \left(\! - \left({\bm{\rho }}
\!+\! {\frac{\sqrt m}{2}} \bf{x}\right)^2 \right) \right)-\]
\begin{equation}
\left.-\!\left(1 \!+\! {\frac{m}{2}}\right)g\exp \left( - \rho ^2
\right)\right\}\!\to - m\left| g \right|C_0  + m\left| g
\right|C_1 x^2 + \ldots\;, \label{E27}
\end{equation}
where ${\bf{x}} \equiv \left( {{\bf{r}}_2  - {\bf{r}}_1 }
\right)/{\sqrt m }$, and all quantities of the dimension of energy
are multiplied by $m$. Then, as $m \to 0$ and for small deviations
from the position of equilibrium, we have the oscillatory
potential ($C_1  > 0$). Since the oscillatory frequency $\omega _0
\sim\sqrt m $, the spectrum becomes denser for smaller masses.
Hence, for constants $ g_{s,\mathrm{cr}}/2 < \left| \lambda
\right| < g_{s,\mathrm{cr}}/{2m}$ and small masses ($m \to 0$), we
have the growing number of bound levels of the equidistant
spectrum. Respectively, the thresholds for small masses will
represent the equidistant spectrum in this region.

In the diagrams of the thresholds of stability (see
Fig.\ref{fig3}), the indicated regularities are manifested, as $m
\to 0$, in the enlargement of ``islands'' (``traps'') of the
infinite series of Efimov levels, the separation (due to the
merging with the earlier isolated thresholds in the places of a
``wedge'') of the increasing number of the new lines of thresholds
from the islands, the ``pulling'' of them into the region of the
``tube'', and the gradual filling of the whole region $\left|
\lambda \right| > g_{s,\mathrm{cr}}/2 $, $g > -
g_{s,\mathrm{cr}}^{(0)}$. It is easy to trace the change of the
general pattern in Fig.\ref{fig3} with decrease in the mass of the
third particle or, conversely, an increase in this mass. We can
qualitatively explain the growth of the number of thresholds in
the ``tube'' by the following reasoning. Because $\Lambda _0 $
depends on the mass of particles \cite{R3} and tends to 1 with
decrease in the mass of the third particle, energy (\ref{E24})
depends more and more weaker on the level number $n$. Therefore,
the number of levels, for which energy (\ref{E24}) remains
negative, increases. That is, the less the mass, the greater is
the number of thresholds which correspond to excited levels and
can be ``pulled'' into the ``tube''. This is completely confirmed
by the high-precision calculations for small masses. In
particular, the first excited level begins to move into the
``tube'' on the transition from great masses to $m\sim 1$
(Fig.\ref{fig1}). In the limit $m \to 0$, the number of levels in
the ``tube'' tends to infinity, but, for any finite mass, the
number of levels is finite (Fig.\ref{fig3}). In the general
aspect, the less the mass of the third particle, the more the
symmetric states become similar to the first excited one (for $m =
1$). This implies that, in a certain sense, the effects of Efimov
and Thomas and the effect of a ``tube'' have many common features.
On the contrary, we may expect that, in the case of the model of
two particles in the field of a fixed center and under a variation
of the forms of interaction potentials, all three effects can be
significantly impoverished, so that they will have few common
features.

\section{Effect of ``Traps'', and Effects of
``Rearrangement'' of Energy Levels}

\begin{figure}[ht]
\centering
\includegraphics[width=6.in]{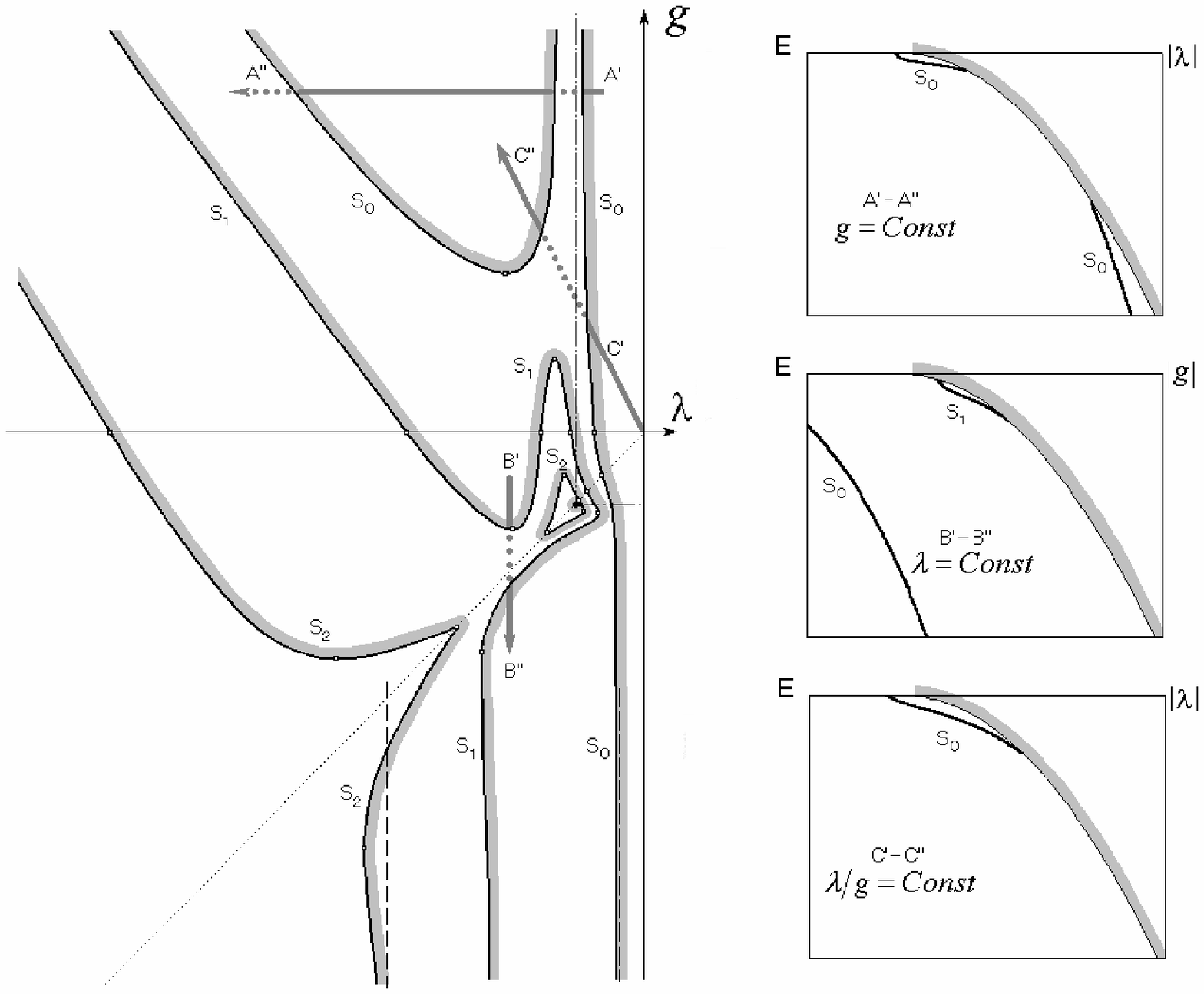}
\caption{To the left, the calculated diagram of the thresholds of
stability (given on an arbitrary scale) is shown, the arrows at
three different places indicate the direction of an increase in
the attraction, the other notations are analogous to those in
Fig.~\ref{fig1}. To the right, the dependence of the energies of
three particles on the strengthening of the attraction is shown:
$A^{'}-A^{''}$ corresponds to the dependence of the ground state
energy on the constant $\lambda$ (at $g=\mathrm{const}$) with the
available ``trap'' in the resonance region, $B^{'}-B^{''}$
corresponds to the dependence of the energies of the ground and
first excited levels on the constant $-g$ (at
$\lambda=\mathrm{const}$) with the available ``trap'' for the
excited state, $C^{'}-C^{''}$ demonstrates the presence of a
``trap'' for the ground state (at $\lambda/g=\mathrm{const}$).}
\label{fig4}
\end{figure}

The nonmonotonous change of the number of levels of the
three-particle system with increase in the interaction constants
in the region of attraction, where the levels appear and then
disappear on a two-particle threshold with  increase in the
attraction (the effect of a ``trap'') is characteristic (see the
diagrams of thresholds in Figs.\ref{fig1} and \ref{fig3}) of a
wide region of the interaction constants which can be positioned
even sufficiently far from the Efimov region of the resonance
interaction. If the general constant of attraction in the
three-particle system grows so that $g$ and $\lambda$ are linearly
related to each other, we can easily reveal a nonmonotonous
character of the behavior (appearance and disappearance) of
levels. That is, the effect of a ``trap'' has a sufficiently
universal character. In Fig.\ref{fig4}, the lines with arrows show
some directions of the coordinated increase in the intensities of
the interaction potentials, where the three-particle levels appear
and then disappear with increase in the interaction constants (the
continuous and dotted lines correspond, respectively, to the
absence and presence of the given bound state). In this figure, we
present (schematically) the results of high-precision calculations
with the Gaussian potentials of unit radii and $m = 1$ (but, for
the convenience of a perception, we choose the scale to be
arbitrary). The figure demonstrates the presence of a ``trap''
even in a very narrow interval of the constant $\lambda $ near a
critical value of the resonance interaction constant $\lambda
\approx  - g_{s,\mathrm{cr}} $, though this region is small. For
the sake of convenience, we schematically present (to the right in
Fig.\ref{fig4}) the dependence of the binding energies on the
coupling constants with strengthening the attraction along the
mentioned directions. In this case, we see the appearance of
``traps'' for the energy levels in both the ground and excited
states near two-particle thresholds. In the general aspect, the
effect of a ``trap'' for three-particle levels, which demonstrates
the essentially different mode of the dependence of the
ground-state energy of two particles and that of the
three-particle levels on the interaction constants, is a
consequence of the two-particle structure of the full interaction
potential in the three-particle system.

\begin{figure}[ht]
\centering
\includegraphics[width=6.in]{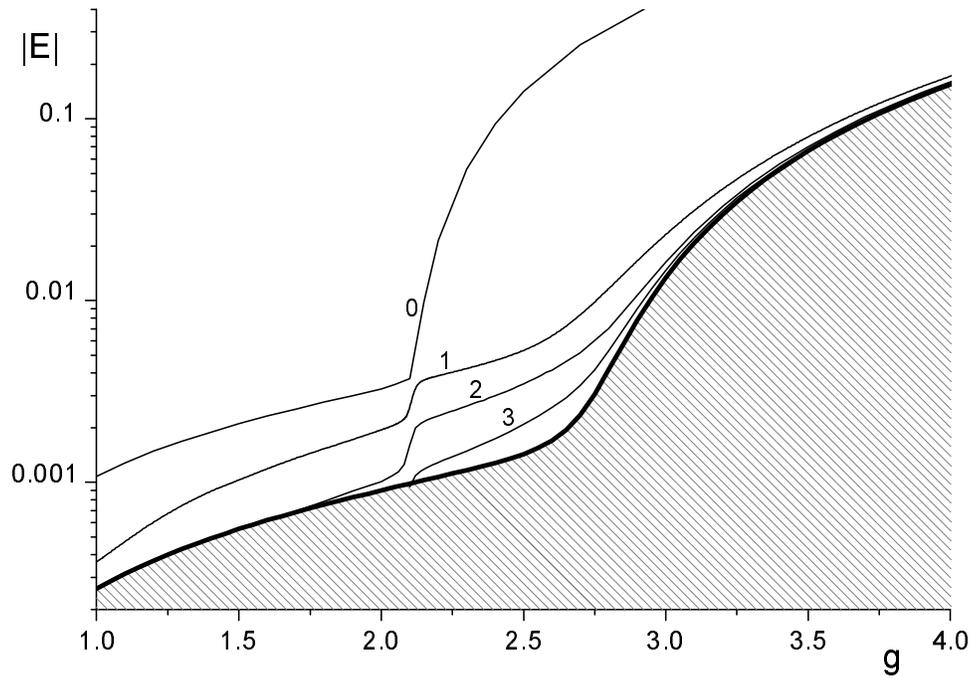}
\caption{ Dependence of the binding energies of three particles on
the coupling constant for the two-component potential with
different radii (\ref{E28}); 0, 1, 2, 3 are the numbers of the
ground and excited levels. The shaded region corresponds to the
continuous spectrum for three particles.} \label{fig5}
\end{figure}

Quite nontrivial is the possibility to vary the modes of behavior
of the thresholds (and of the energy levels) with the help of the
interaction potentials which involve at least two modes of
attraction with essentially different radii. In this case, we can
nontrivially realize a certain generalization of the well-known
Zel'dovich two-particle effect of a ``rearrangement'' of the
energy spectra in the system of three particles. For the pairwise
interaction potentials with the components of different radii, for
example, of the type
\begin{equation}
U(r) = V(r) =  - g\left( {\exp \left( { - r^2 } \right) + b\exp
\left( { - \left( r/ {r_0 } \right)^2 } \right)} \right)
\label{E28}
\end{equation}
(where $r_0\gg 1$ and $b>0$), one observes a ``rearran\-gement''
(Fig.\ref{fig5}) of the energy spectrum in the three-particle
system (a three-particle analog of the Zel'dovich effect). Due to
a great value of the ratio of the interaction radii of the
two-component attractive potential ($r_0 \gg 1$), the spectrum of
three particles is close to the superposition of the spectra of
two problems involving three particles with the separately taken
components of the interaction potential, and the quasidegeneration
of the levels occurs. But, at the points of the expected
intersection of levels of these two spectra, we observe the
effects related to the ``rearrangement'' of levels.  For example,
the first excited level replaces the ground-state level with
increase in the total attraction (with increase in $g$). In its
turn, the ground-state level sharply changes the mode of the
dependence on the interaction constant. In Fig.\ref{fig5}, on the
real scale, we present the results of calculations of a
three-particle completely symmetric energy spectrum (the binding
energies) as a function of the interaction constant $g$ for a
specific example of potential (\ref{E28}) ($m = 1,$ $b = 0.001,$
$r_0  = 100$). At the same interaction constant $g \approx 2.1$,
the ``rearrangement'' of the spectrum of four levels occurs
successively: $n$-th level replaces $(n + 1)$-th level. The ground
state level is the only one which, after the ``rearrangement'',
corresponds to the binding energy with a quite different mode of
the dependence on $g$. The effect of a ``rearrangement'' of the
three-particle spectrum is much more pronounced than that for the
two-particle threshold and occurs for a less attraction. By
changing the class of the corresponding interaction potentials and
their parameters, we can vary the number of levels in the region
of a ``rearrangement'' and can also create a ``trap'' for other
levels (see results in \cite{R7} for nonlocal separable
potentials).  The close conditions for a realization of the
effects of ``rearrangement'' and ``traps'' for the energy states
are known in solid state physics for a long time in the case
where, for example, the conditions for a ``quasidegeneration'' of
the phonon and optical branches of oscillations are realized, or
the crystals with admixtures are considered.

\begin{figure}[ht]
\centering
\includegraphics[width=6.in]{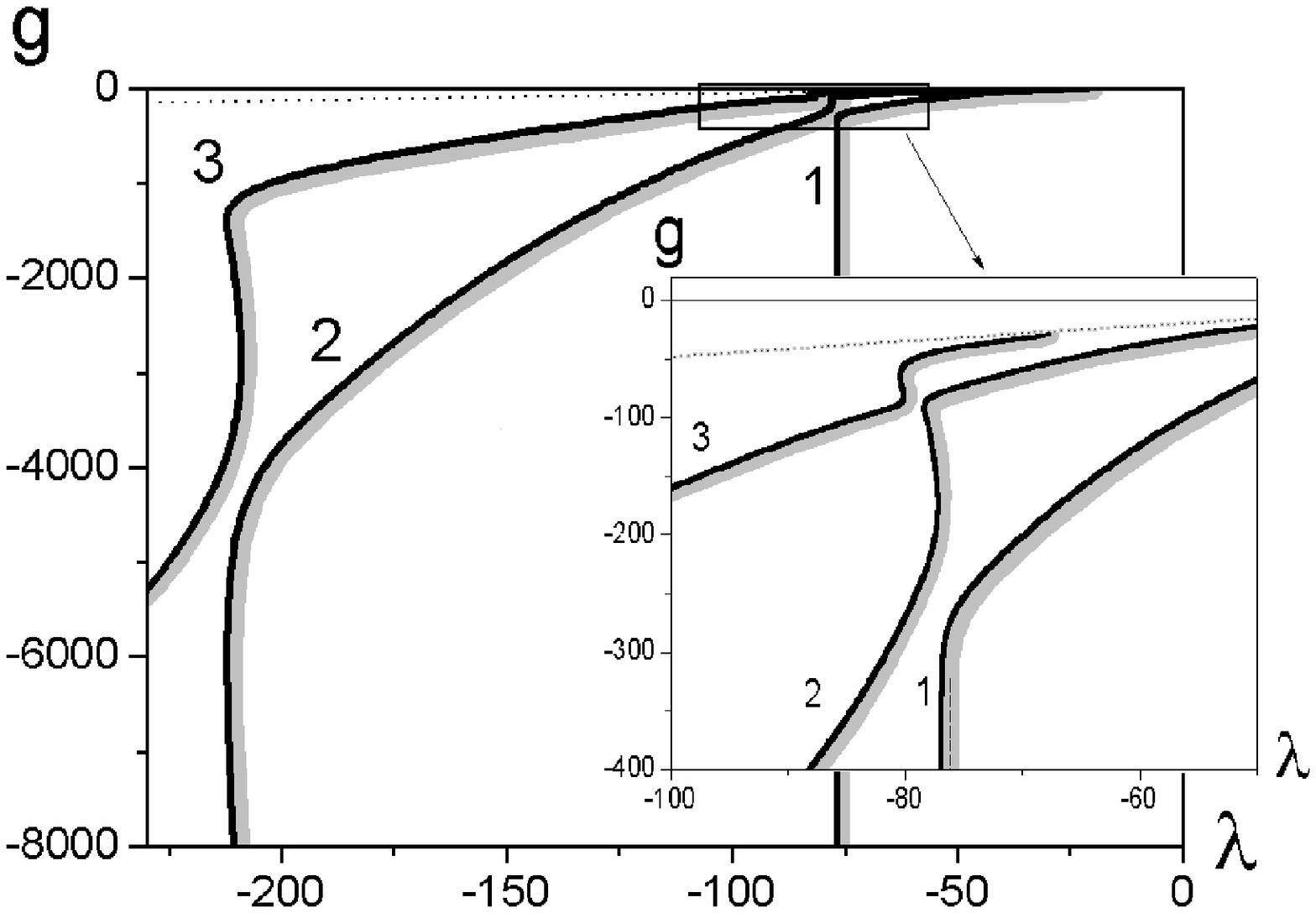}
\caption{Calculated diagram with the ``rearrangement'' of the
lines of the thresholds of stability (on the real scale) for the
second and third excited levels at $m=0.06$ and for the Gaussian
potential. The insert shows the ``rearrangement'' effect of
thresholds on a greater scale.  Dots show the line of the
two-particle threshol.} \label{fig6}
\end{figure}

Even for simple potentials in the Gaussian form, the performed
high-precision studies of the energy thresholds for three
particles reveal one more effect of a sharp change of the modes in
the diagrams of the thresholds of stability which can be also
named the effect of a ``rearrangement''. This effect has no analog
on the two-particle level. There exist the regions of parameters
(in the $\lambda $ axis, these are the regions close to the
vertical asymptotes in Fig.\ref{fig3}), where the
``rearrangement'' of the adjacent energy thresholds occurs in the
digram of thresholds. That is, on the monotonous change of the
interaction intensities, the threshold of the $(n+1)$-th level
sharply approaches that of the $n$-th one and occupies its place.
This effect is especially clearly manifested in the case of small
masses of the third particle (see Fig.\ref{fig6}, where the lower
part of the diagram of the thresholds of three first excited
levels calculated for the mass $m = 0.06$ is shown). We note that,
in Fig.\ref{fig3}, this effect of the rearrangement of thresholds
is not marked in order to avoid the complication of a perception
of the general scheme of three-particle thresholds. The presence
of the effect of a ``rearrangement'' is probably related to the
following. In addition to the considerable short-range attractive
interaction between particles ${\it (12)}$ present in Hamiltonian
(\ref{E1}), there appears an additional effective oscillatory
interaction having a great radius at small masses of the third
particle. This additional interaction between particles ${\it
(12)}$ is generated by the motion of the third light particle.
Then, as $m \to 0$, the radius of the oscillatory well grows. As
known\cite {R7}, in the presence of two potentials with
essentially different radii, there occurs the ``rearrangement'' of
the energy spectrum on increasing the intensities of the
potentials. In our case where the intensity of the short-range
attraction becomes close to the critical constant of the
appearance of the $n$-th level in this well, the lowest level of
the other well falls into the well with less radius. On its
previous place, the next level falls, etc. In this case, the
spectrum for the well with a greater radius is practically renewed
and is preserved, until the increase of the intensity of the
short-range attraction induces the appearance of a new, $(n +
1)$-th level, and the ``rearrangement'' repeats. As seen from
Fig.\ref{fig6}, this region of the rearrangement of thresholds
contains also ``traps'', because the lines of thresholds depending
on $g$ are not monotonous.

Finally, we note that, for the three-particle problem, the
nonmonotonicity of the thresholds of stability is a general
regularity, which is rather a rule than the exception. One can
assume that the nonmonotonicity of thresholds in a three-particle
system is a consequence of the difference of the modes of
dependence of the ground-state energy of two particles and
three-particle energies for the chosen, even simple, interaction
potentials. Indeed, for simple attractive potentials in a
two-particle system, one mode is always realized; we will say that
this is one configuration. At the same time, for the system of
three particles with pairwise interaction potentials and greater
number of coordinates, various spatial configurations of three
particles can be realized. This can induce a change of the modes
of the dependence of three-particle energies and thresholds and,
therefore, the nonmonotonous behavior of the number of levels of
three particles in various regions of parameters of the
interaction potentials. We may expect a more complicated behavior
of the thresholds of stability in the problems with four and
greater number of particles with pairwise interaction potentials.

\section{Conclusions}

In conclusion, we emphasize that we have systematically studied
the energy thresholds of stability for the general quantum problem
of three particles interacting by short-range potentials.
Analytically and by the high-precision calculations of the energy
spectra of three particles, we have established the general
nonmonotonic dependence of the thresholds of stability on the
strengthening (or the weakening) of the attraction between
particles. As a result, we have revealed a number of new
nontrivial universal effects in the behavior of the three-particle
thresholds of stability. In the first turn, we mark the effect of
a ``trap'', when the number of three-particle bound states
nonmonotonically depends on the strengthening of the interaction
between particles. Secondly, we have discovered the effect of a
``rearrangement'' of the energy levels and the thresholds of
stability, which has no analog in the two-particle systems. At the
same time, we have constructed a three-particle generalization of
the Zel'dovich effect of a ``rearrangement'' of the energy levels
for the attractive interaction potentials with the modes of
attraction which essentially differ by their interaction radii.
Thirdly, we have revealed both the effect of a ``tube'' for the
energy states in the region of a significant repulsion between a
pair of identical particles and the effect of a ``wedge'' on the
boundary, where the binding energies of two different pairs of
particles coincide. We have made a sufficiently full and clear
analysis of the dependences of the thresholds of stability on the
masses of particles and the form of a short-range interaction, and
have also considered different symmetries of states.

The revealed regularities are important for the comprehension of
the general properties of three-particle systems of different
nature. In addition, the possibilities to describe the
characteristic peculiarities of thresholds in three-particle
systems with the Coulomb interaction, as well as those of the
states with nonzero angular momenta, are open. These problems will
be analyzed in further publications.  Challenging is also the
necessity to study the thresholds of stability for one- and
two-dimensional systems, where one can expect significant
differences in properties from those in three-dimensional
problems.

\clearpage

\clearpage

\clearpage

\clearpage

\clearpage

\clearpage

\begin{thebibliography}{9}

\bibitem{R1} E. Nielsen, D.V. Fedorov, A.S. Jensen, E. Garrido,
Phys. Rep. {\bf 347}, N 5, 373---459 (2001).
\bibitem{R2} V.N. Efimov, Yad. Fiz. {\bf 12}, Iss. 5, 1080---1091 (1970).
\bibitem{R3} B.E. Grinyuk, M.V. Kuzmenko, I.V. Simenog,
Ukr. Fiz. Zh. {\bf 48}, N 10, 1014---1023 (2003), LANL e-print
nucl-th/0308003.
\bibitem{R4} M.Kh. Shermatov, Teor. Mat. Fiz. {\bf 136}, N~2, 257---270 (2003).
\bibitem{R5} L.H. Thomas, Phys. Rev. {\bf 47}, N~12, 903---909 (1935).
\bibitem{R6} B.E. Grinyuk, I.V. Simenog, in: {\it Physics of
Multiparticle Systems} (Naukova Dumka, Kyiv, 1985 (in Russian)),
Iss. 8, 66---95.
\bibitem{R7} I.V. Simenog, A.I. Sitnichenko,
Ukr. Fiz. Zh. {\bf 28}, N 1, 1---6 (1983).

\end{thebibliography}
\end{document}